\documentclass[aps,prl,twocolumn ,groupedaddress,longbibliography]{revtex4-1}
\usepackage{graphicx}
\usepackage{amsmath}
\usepackage{float}
\usepackage{hyperref}
\usepackage{color}
\usepackage[british]{babel}

\begin{document}


\title{Cooling and state preparation in an optical lattice via Markovian feedback control}


\author{Ling-Na Wu}
\email[]{lingna.wu@tu-berlin.de}
\affiliation{Institut f\"ur Theoretische Physik, Technische Universit\"at Berlin,
	Hardenbergstra\ss e 36, Berlin 10623, Germany}
\author{Andr{\'e} Eckardt}
\email[]{eckardt@tu-berlin.de}
\affiliation{Institut f\"ur Theoretische Physik, Technische Universit\"at Berlin,
	Hardenbergstra\ss e 36, Berlin 10623, Germany}


\date{\today}

\begin{abstract}
	We propose and investigate a scheme based on Markovian feedback control that allows for the preparation of single targeted eigenstates of a system of bosonic atoms in a one-dimensional optical lattice with high fidelity. It can be used for in-situ cooling the interacting system without particle loss, both for weak and strong interactions, and for experimentally preparing and probing  individual excited eigenstates. For that purpose the system is assumed to be probed weakly via homodyne detection of photons that are scattered off-resonantly by the atoms from a structured probe beam into a cavity mode. By applying an inertial force to the system that is proportional to the measured signal, the system is then guided into a pure target state. 
	The scheme is found to be robust against reduced measurement efficiencies.
\end{abstract}


\maketitle

\textit{Introduction.}---
Atomic quantum gases in optical lattices constitute a unique experimental platform
for studying quantum many-body systems under extremely clean and
flexible conditions~\cite{lewenstein2012ultracold}.
An important property of these systems is their extremely high degree
of isolation from the environment, which is provided by the optical trapping of
atoms inside ultrahigh vacuum. It offers the rather unique opportunity to
experimentally study coherent nonequilibrium dynamics of many-body 
systems~\cite{RevModPhys.83.863} over
hundreds of tunneling times, before eventually residual heating
processes make themselves felt.
However, suitable dissipation can also be beneficial and the lack of it a limitation. For instance, 
whenever excitations are created, e.g.\ by time-dependent parameter variations into a parameter regime of interest,
the induced energy remains in the system and causes heating. Here the ability to cool 
the system after a parameter ramp \emph{in situ} (without particle loss) would be highly desirable. Also the controlled preparation of states beyond the strict constraints of equilibrium in driven-dissipative systems offers intriguing perspectives, like the investigation of transport~\cite{Landi2021,Bertini2021}, the engineering of non-hermitian Hamiltonians~\cite{Bergoltz2021},  Bose condensation in non-equilibrium steady states~\cite{Vorberg2013, Schnell2017}, etc.. Therefore, various routes for engineering dissipation in quantum-gas experiments have been pursued already, such as the engineering of 
reservoirs~\cite{Krinner2017} and heat baths~\cite{Schmidt2018}, the implementation of dephasing noise via inelastic scattering of lattice photons~\cite{Pichler2010}, or engineered local particle loss/detection via ionization~\cite{Ott2016}. 

Here we propose and investigate a scheme for controlled dissipative state preparation and cooling in bosonic optical-lattice 
systems by means of measurement-based feedback control~\cite{wiseman2009quantum,zhang2017quantum}.
It is based on the continuous measurement of photons that are off-resonantly scattered by the atoms from a structured probe beam into a cavity mode \cite{RevModPhys.85.553, PhysRevLett.114.113604} in combination with a simple lattice measurement-conditioned acceleration and can be used to prepare individual eigenstates of the system with high fidelity. Controlling the probe beam as well as the feedback strength allows for engineering an artificial environment with tailored properties. It permits preparing the ground state (i.e.\ cooling) of the system with high fidelity not only for non or weakly interacting bosons, but, remarkably, at integer filling also for arbitrarily strong repulsion. Moreover, the approach is capable also of preparing individual excited eigenstates of the system, which would open the possibility to study their statistical properties as they are predicted, e.g., for ergodic systems by  the eigenstate thermalization hypothesis (ETH)~\cite{Rigol2008,kaufman2016quantum,d2016quantum}. 

Measurement-based feedback control has been proposed and used already successfully in other systems for quantum state preparation~\cite{PhysRevLett.99.223601,PhysRevLett.115.060401,PhysRevA.94.052120,PhysRevLett.97.073601,PhysRevLett.97.190201,Sayrin2011,PhysRevLett.108.243602,Geremia270,PhysRevLett.110.163602,PhysRevLett.116.093602,PhysRevLett.117.073604,Riste2013,PhysRevX.7.011001},
cooling~\cite{PhysRevA.95.043641,Hush_2013,PhysRevA.80.013614,PhysRevA.75.051405,PhysRevA.100.063819,Ivanov_2014,PhysRevLett.96.043003,PhysRevLett.105.173003,PhysRevLett.111.103601,PhysRevLett.105.173003,Wilson2015},
simulating nonlinear dynamics~\cite{PhysRevA.62.012307,PhysRevLett.124.110503,PhysRevA.102.022610},
as well as controlling dynamics~\cite{PhysRevLett.110.013601,Vijay2012,PhysRevLett.88.093003,PhysRevLett.92.223004,PhysRevLett.110.210503,Kroeger_2020}
and phase transitions~\cite{Kopylov_2015,PhysRevLett.124.010603,PhysRevResearch.2.043325,PhysRevA.99.053612,2021arXiv210202719B}, just to name a few. In this paper, we consider Markovian feedback control as proposed by Wiseman~\cite{PhysRevA.49.2133}, where a feedback Hamiltonian proportional to the measurement signal is continuously added to the system. It has been employed already for the stabilization of arbitrary one-qubit quantum states~\cite{PhysRevA.64.063810,PhysRevLett.117.060502}, the control of two-qubit entanglement~\cite{PhysRevA.71.042309,PhysRevA.76.010301,PhysRevA.78.012334,Wang2010}, as well as optical and spin squeezing~\cite{PhysRevA.50.4253,PhysRevA.65.061801,2021arXiv210202719B}.


\textit{Model.}---
{We consider a system of $N$ interacting bosonic atoms in a one-dimensional optical lattice described by the} Bose-Hubbard model,
\begin{equation}\label{BH}
	H = -J\sum\limits_{l=1}^{M-1}(a_l^\dag a_{l+1} + a_{l+1}^\dag a_l) + \frac{U}{2}\sum\limits_{l=1}^{M}n_l (n_l-1),
\end{equation}
with annihilation and number operators $a_l$ and $n_l=a_l^\dag a_l$ for bosons on site $l$. Here $J$ quantifies nearest-neighbor tunneling and $U$ on-site interactions.

In order to control the system via quantum feedback control, we consider a homodyne measurement of an operator $c$.
The dynamical evolution of the system
is then described by the stochastic master equation~(SME)~\cite{wiseman2009quantum}~($\hbar=1$ hereafter),
$d\rho_c = -i[H,\rho_c] dt + {\cal D}[c]\rho_c dt + {\cal H}[c]\rho_c dW$,
with ${\cal H}[c]\rho := c\rho + \rho c^\dag - {\rm Tr}[(c+c^\dag)\rho]\rho$
and ${\cal D}[c]\rho := c\rho c^\dag - \frac{1}{2}(c^\dag c \rho + \rho c^\dag c)$.
Here $\rho_c$ denotes the quantum state conditioned on the measurement result,
$I_{\rm hom} = {\rm Tr}[(c+c^\dag)\rho] + \xi(t)$, with $\xi(t)=dW/dt$ and $dW$
being the standard Wiener increment with mean zero and variance $dt$.
The quantum backaction of a weak measurement can be used for tailoring the system's dynamics and state. For instance, a quantum nondemolition measurement of light
allows for the preparation of different types of
atom-number squeezed and macroscopic superposition
states~\cite{PhysRevLett.102.020403,PhysRevLett.114.113604}.
Even more control is obtained by introducing measurement-dependent feedback. 

Here we consider Markovian feedback control, where a feedback term
$I_{\rm hom} F$ proportional to the instantaneous signal is added to the Hamiltonian, giving rise to an evolution governed by the modified SME~\cite{PhysRevA.49.2133}
$ d\rho_c = -i[H+H_{\rm fb},\rho_c]dt + {\cal D}[A]\rho_c dt +{\cal H}[A]\rho_c dW$,
with operators
\begin{equation}\label{A}
	A = c-iF,  \quad H_{\rm fb}=\frac{1}{2}(c^\dag F+F c).
\end{equation}
Taking the ensemble average over the stochastic measurement outcomes gives the feedback-modified ME~\cite{PhysRevA.49.2133}
\begin{equation}\label{me_fbM}
	\dot \rho = -i[H+H_{\rm fb},\rho] + {\cal D}[A]\rho.
\end{equation}
The effect induced by the feedback loop is seen to replace the collapse
operator $c$ by $A$ and to add an extra term $H_{\rm fb}$ to the Hamiltonian.
Our goal is to find the proper measurement operator $c$ and feedback operator $F$ so that the
steady state of the ME~\eqref{me_fbM} is (as close as possible to) an eigenstate $|E\rangle $ of the Hamiltonian $H$ \eqref{BH}.
Ref.~\cite{PhysRevA.72.024104}
points out that the ME~\eqref{me_fbM} has a pure steady state if
and only if the effective Hamiltonian
$H_{\rm eff} = H + H_{\rm fb} -iA^\dag A/2$ and the collapse operator $A$ have a common eigenstate, which is then the steady state.
Since the additional feedback-induced term in the Hamiltonian $H_{\rm fb}$ is proportional to the measurement strength,
for weak measurement, we can safely neglect $H_{\rm fb}$ relative to $H$ and construct a collapse operator $A$ so that $A|E\rangle = 0$.
The dissipator cannot be neglected, since it is the only dissipative term.
This strategy is confirmed in our simulations below, where $H_{\rm fb}$ is fully taken into account. In the following we will first discuss the preparation of the ground state $|G\rangle$, before considering also excited eigenstates.

\textit{Feedback scheme for non-interacting case.}---
In order to get an insight of how to choose the measurement and feedback operator,
let us first take a look at the double-well system~($M=2$).
It can be mapped to a spin-$N/2$ system~\cite{dalton2012two} with
the collective spin operators defined as $J_x = \frac{1}{2}(a_1^\dag a_2 + a_2^\dag a_1)$,
$J_y = -\frac{i}{2}(a_1^\dag a_2 - a_2^\dag a_1)$, and $J_z = \frac{1}{2}(a_1^\dag a_1 - a_2^\dag a_2)$,
which are angular momentum operators satisfying the commutation relations
$[J_\mu,J_\nu]=i\epsilon_{\mu\nu\sigma}J_\sigma$ for $\mu,\nu,\sigma=x,y,z$ and
the Levi-Civita symbol $\epsilon_{\mu\nu\sigma}$.
The ground state of the non-interacting Hamiltonian $H_2 = -J(a_1^\dag a_2 + a_2^\dag a_1) = -2JJ_x$
is the eigenstate of $J_x$, $|j=N/2,m=N/2\rangle$, with the largest eigenvalue
$m=N/2$. 
From the theory of angular momentum, we know that this state
is also an eigenstate of operator $J_-=J_z-iJ_y$ with eigenvalue $0$. 
Hence,
by choosing the measurement operator $c$ as $\sqrt{\gamma}J_z$ and feedback operator $F$ as
$\sqrt{\gamma}J_y$, so that $A=\sqrt{\gamma}(J_z-iJ_y)$, we will reach a steady state very close to the ground state of the
Hamiltonian $H_2$ for weak measurement with strength $\gamma \ll J$.

Generalizing this scenario to the multi-site case, we consider the measurement and feedback operator
\begin{equation}\label{cF}
	c = \sqrt{\gamma}\sum_{l=1}^{M}{z_l n_l}, \quad F = \sqrt{\gamma}\sum_{l=l}^{M-1}(ia_{l+1}^\dag a_{l} - i a_{l}^\dag a_{l+1}).
\end{equation}
Here $c$ is a weighted sum of site occupations and can be implemented via homodyne detection of
the off-resonant scattering of structured probe light from the atoms~\cite{PhysRevLett.114.113604,PhysRevLett.115.095301,RevModPhys.85.553}.
{The detection efficiency can be enhanced by placing the system inside an optical cavity~\cite{Brennecke2007,landig2015measuring,Kroeger_2020,RevModPhys.85.553,2021arXiv210204473M}, so that via the Purcell effect~\cite{purcell1995spontaneous} the photons are scattered predominantly into one cavity mode.}
The feedback $F$ is given by an imaginary component of the  tunneling parameter and can be realized by accelerating the lattice, as discussed below.
For the non-interacting case~($U=0$), the ground state of~\eqref{BH} reads $|G\rangle =\frac{1}{\sqrt{N!}}\big( \sum\nolimits_{l}{g_l}a_l^\dag\big)^N|0\rangle$ with $g_l = \sqrt{\frac{2}{M+1}} \sin\big(\frac{\pi l}{M+1}\big)$.
The condition $A|G\rangle =(c-iF)|G\rangle= 0$ then fixes the coefficients in the
measurement operator 
to $z_l = ({g_{l+1} - g_{l-1}})/{g_l}.$

\begin{figure}
	\centering
	\includegraphics[width=0.99\columnwidth]{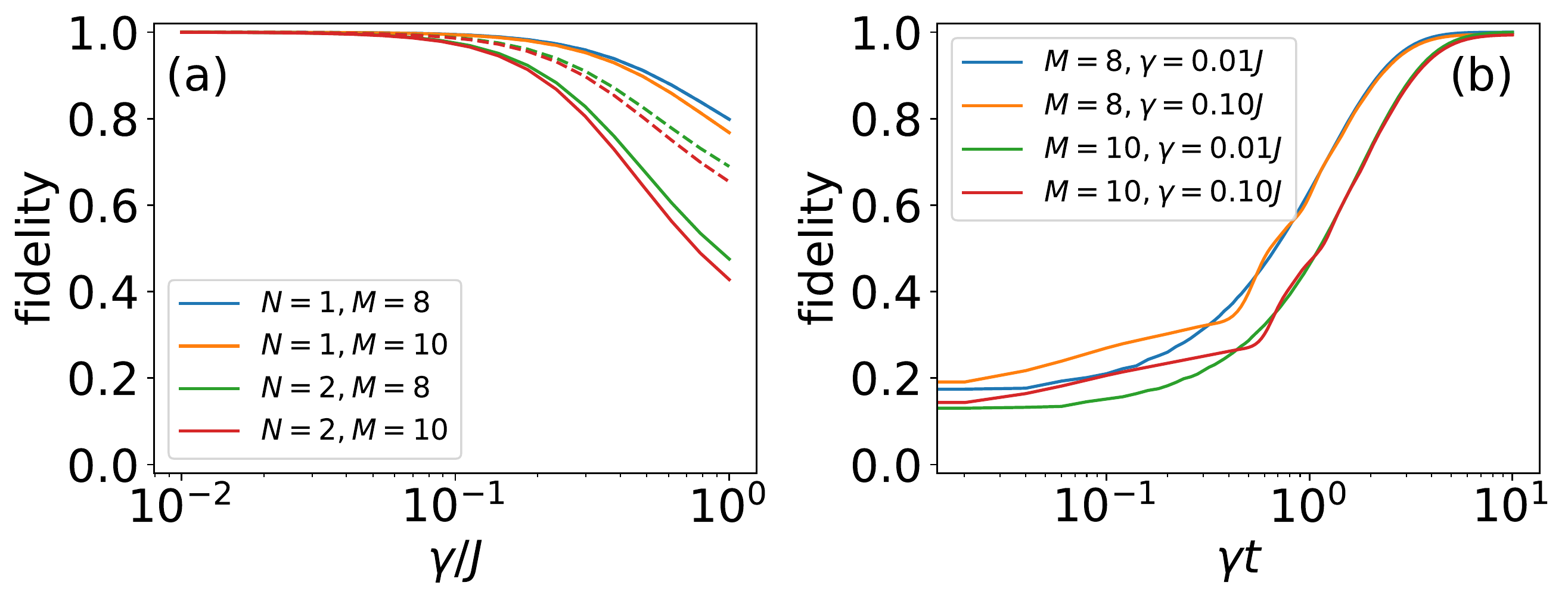}
	\caption{ 
		(a)~Fidelities $f$~(solid) and $\tilde f$~(dashed) between the steady state of Eq.~\eqref{me_fbM} and
		the ground state of Eq.~\eqref{BH} versus measurement strength~$\gamma$.
		(b)~Fidelity between the instantaneous state and the ground state of Eq.~\eqref{BH}, starting from the state with one particle on the leftmost site.}\label{fig0}
\end{figure}

In order to check the proposed scheme, 
we calculate the fidelity $f$ between the steady state $\rho_{\rm ss}$ of Eq.~\eqref{me_fbM} and the ground state $|G\rangle$ of Eq.~\eqref{BH}, i.e., $f = \sqrt{\langle G| \rho_{\rm ss} |G \rangle}$,
which takes values between $0$ and $1$. To compare results for different particle numbers, we also introduce the fidelity ``per particle'' $\tilde f \equiv f^{1/N}$.
Figure~\ref{fig0}(a) shows the fidelity as a function of the measurement strength $\gamma$. 
As expected, for small enough $\gamma$, $H_{\rm fb}$ has no detrimental impact and
a close-to-one fidelity is approached~{\footnote{For odd site number $M$, the steady state is not unique due to degeneracies in the spectrum. However, by adding tiny interactions  a unique steady state is recovered.}}.
This is confirmed when considering the excitation energy $\Delta E={\rm Tr}[\rho_{\rm ss}H]-\langle G|H|G\rangle$ as a probe of deviations from the target state~\cite{sm}. In the presence of an external trapping potential, the scheme can directly be adapted, simply by replacing the coefficients $g_l$  by the discrete single-particle ground-state wavefunction of trapped system and even without being adapted it is robust against weak potential modulations~(see~\cite{sm} for more details). 

The coupling $\gamma$ also influences the time required to reach the steady state.
For one particle in a double-well system~($N=1$, $M=2$), it is easy to verify that $f^2=1-\frac{1}{2}e^{-4\gamma t}$~\cite{sm}.
Hence, a close-to-one fidelity can be reached at $\gamma t \gtrsim 1$.
In Fig.~\ref{fig0}(b), we show the time-dependent fidelity $f= \sqrt{\langle G| \rho_{\rm}(t) |G \rangle}$ as a function of the scaled time $\gamma t$ for $N=1$. For $\gamma t \gtrsim 1$ the curves for different $\gamma$ collapse and quickly approach the steady state. From Fig.~\ref{fig0} we can infer that $\gamma \lesssim 0.1 J$ is sufficient
to achieve a high fidelity, corresponding to a preparation in a few tens of tunnelling times.
For  a larger system~(with larger $M$ or $N$) smaller $\gamma$ and thus also longer times are required~\cite{sm}.

\begin{figure}
	\centering
	\includegraphics[width=0.49\columnwidth]{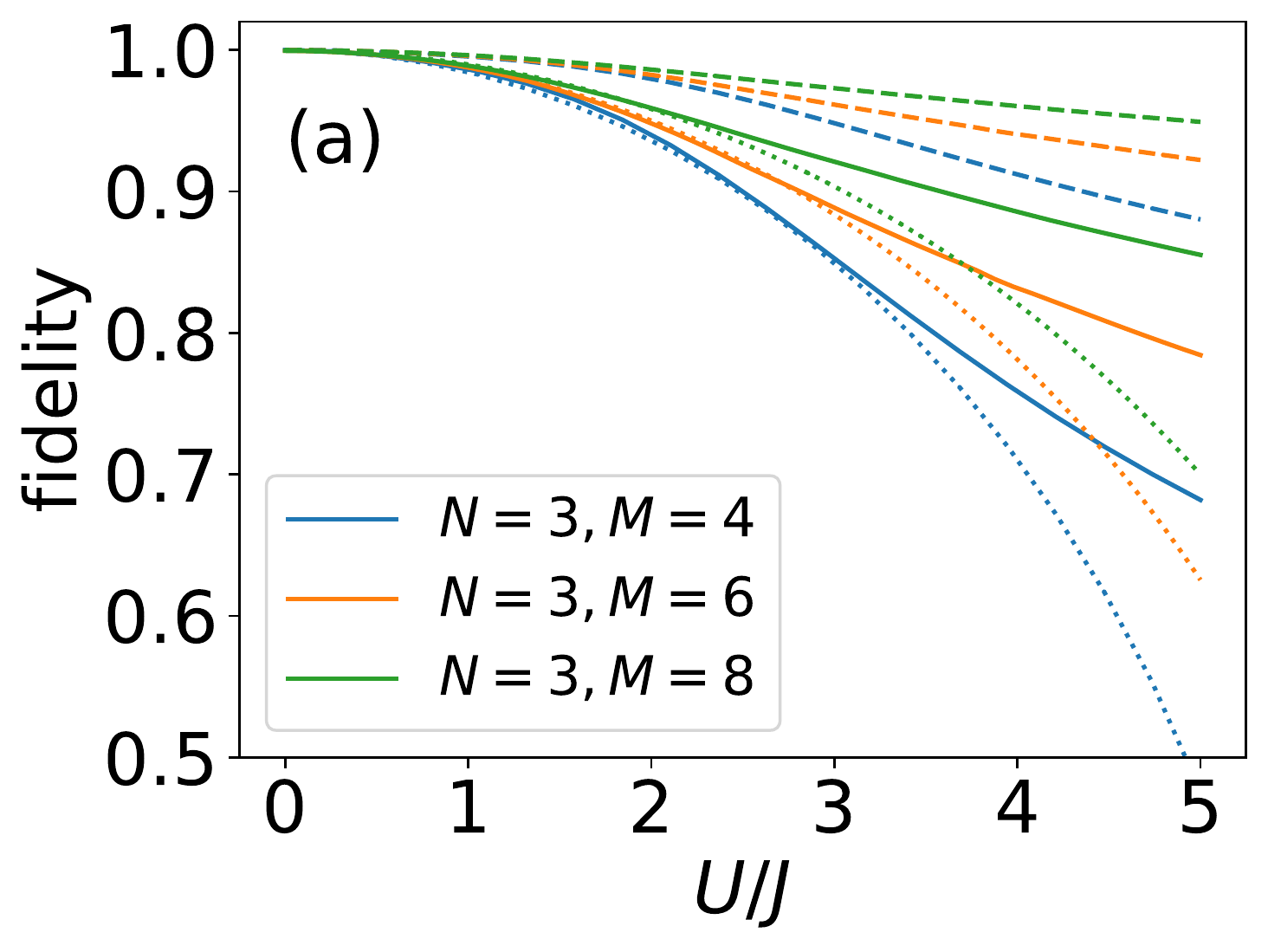}
	\includegraphics[width=0.49\columnwidth]{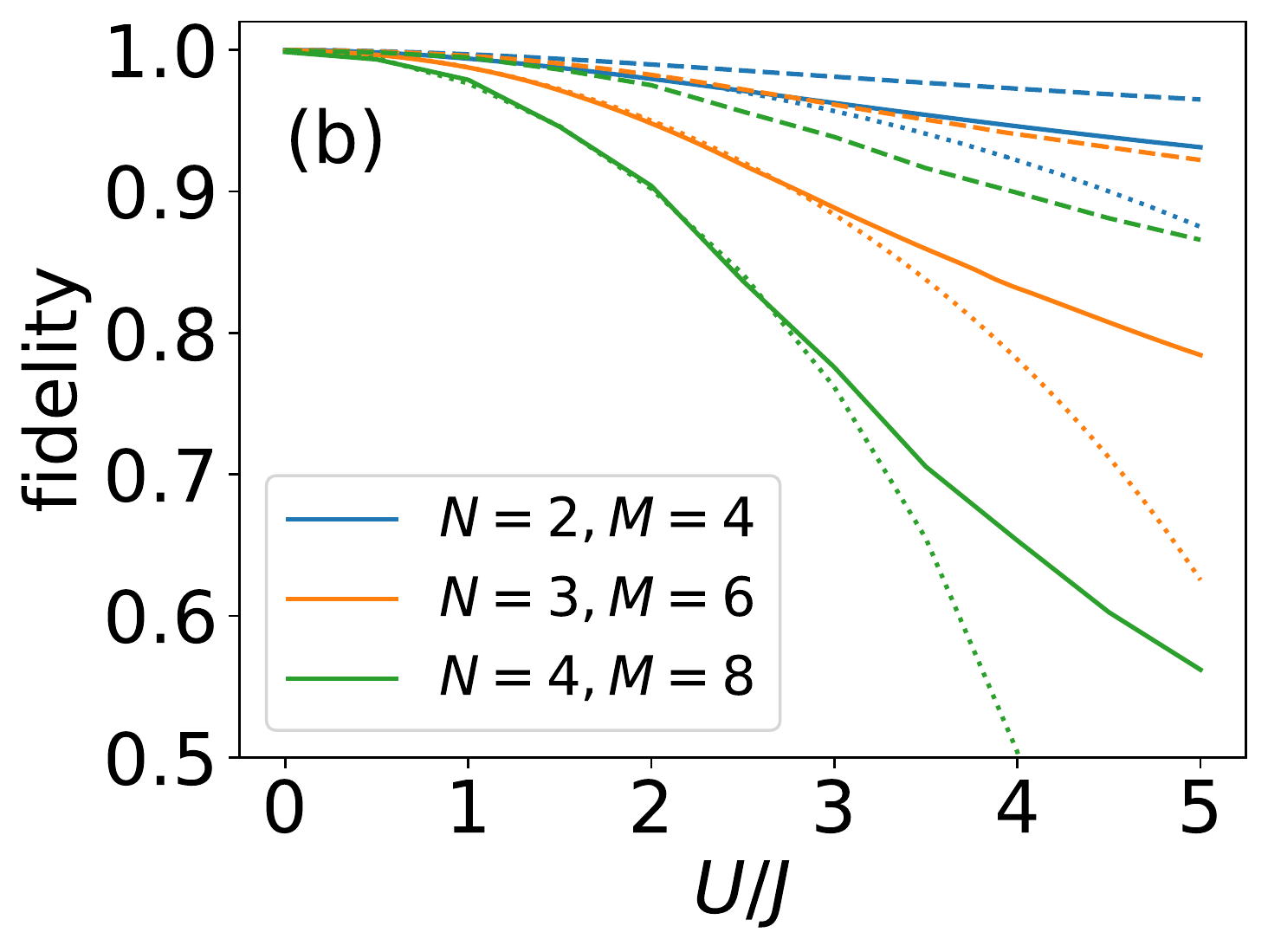}\\
	\caption{Steady state fidelities $f$~(solid) and $\tilde f$~(dashed) versus interaction strength $U$ for different site numbers $M$ at (a) $N=3$ and (b) half filling $N=M/2$, with $\gamma=0.01J$. The dotted lines denote the fitting curves $f^2=1-\chi[(nU)/(2\Delta E)]^2$ with $\chi$ being the fitting parameter~\cite{sm}.}\label{U}
\end{figure}

Without any modification the feedback scheme still performs well in the presence of weak interactions $U\lesssim J$, as can be seen from Fig.~\ref{U} showing the steady-state fidelity  
versus $U$. While for fixed total particle number $N$ the scheme works better for larger system sizes~($M$)~[Fig.~\ref{U}(a) for $N=3$], 
the opposite trend is observed when instead fixing the filling $n=N/M$~[see Fig.~\ref{U}(b) for $n=1/2$].

\begin{figure}
	\centering
	\includegraphics[width=0.95\columnwidth]{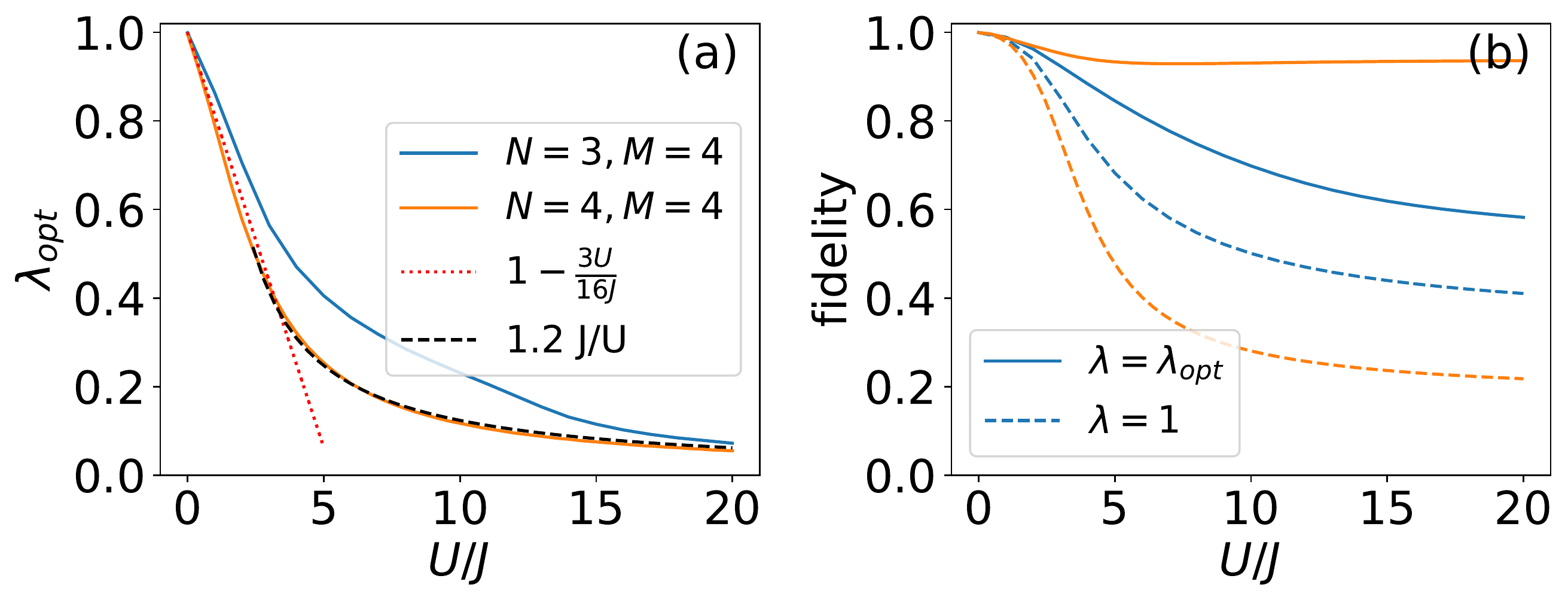}
	\caption{
		(a) The optimal parameter $\lambda$ which gives the highest fidelity between the steady state of Eq.~\eqref{me_fbM} and the ground state of Eq.~\eqref{BH} as a function of the interaction strength $U$. The corresponding fidelity $f$ is shown in (b) 
		for $\gamma=0.01J$.}\label{opt}
\end{figure}

\textit{Feedback scheme for interacting particles.}---
To better capture the impact of interactions, we adapt the feedback scheme by adding a prefactor $\lambda$ to the feedback operator $F$, so that the collapse operator becomes
\begin{equation}\label{cp}
	A= c -i \lambda F,
\end{equation}
The choice of $\lambda$ maximizing the fidelity is shown in Fig.~\ref{opt}(a) as a function of $U$. The corresponding fidelity is shown in (b). We can observe that in the case of integer filling (orange line), this simple rescaling of the feedback strength allows to achieve fidelities close to unity for all interaction strengths. The fact that it is possible to prepare the ground state of an interacting system for every interaction strength, including the strongly correlated regime of intermediate interactions close to the Mott transition \cite{Sachdev2009}, is remarkable, especially since we are considering only single-particle (i.e. quadratic) measurement and feedback operators $c$ and $F$ and since it is sufficient to vary a single parameter $\lambda$ to cover the whole range of interactions.

This choice (\ref{cp}) can again be motivated by inspecting the double-well case.
For weak interactions, the ground state is given by
$|G\rangle \simeq |\psi_0^{(0)}\rangle - \alpha |\psi_2^{(0)}\rangle$
where $|\psi_n^{(0)}\rangle$ denotes the $n$th eigenstate of the non-interacting Hamiltonian and
$\alpha=\sqrt{2N(N-1)}(U/J)/16$.
Since
$A|G\rangle \propto [\sqrt{N}(1-\lambda)-\alpha\sqrt{2(N-1)}(1+\lambda)]|\psi_1^{(0)}\rangle$,
the cooling condition can be satisfied by setting $\lambda = 1-(U/J)(N-1)/(4M)$,
which is a linear function of $U/J$.
For integer filling, the ansatz~\eqref{cp} is also suitable for strong interactions.
The ground state is then approximated by
$|G\rangle \simeq |n, n\rangle + (nJ/U) (|n+1,n-1\rangle + |n-1,n+1\rangle)$ with integer $n=N/2$.
This gives $A|G\rangle = \sqrt{\gamma}[n_1-n_2-\lambda(a_1^\dag a_2 - a_2^\dag a_1)]|G\rangle \simeq \sqrt{\gamma}n\left(2J/U-\lambda\right) (|n+1,n-1\rangle - |n-1,n+1\rangle)$.
Therefore, by setting $\lambda = 2J/U$, we have
$A|G\rangle \simeq 0$, which leads to a steady state close to the ground state of the system.
For half-integer filling (odd $N$), the condition $A|G\rangle \simeq 0$ does not hold for any $\lambda$~\cite{sm}.
Generalizing this reasoning to $M$ sites, we again find $\lambda \propto U/J$ for weak interactions and for integer filling also $\lambda \propto  J/U$ for strong interactions [dashed and dotted lines in Fig.~\ref{opt}(a)]. 

\begin{figure}
	\centering
	\includegraphics[width=0.49\columnwidth]{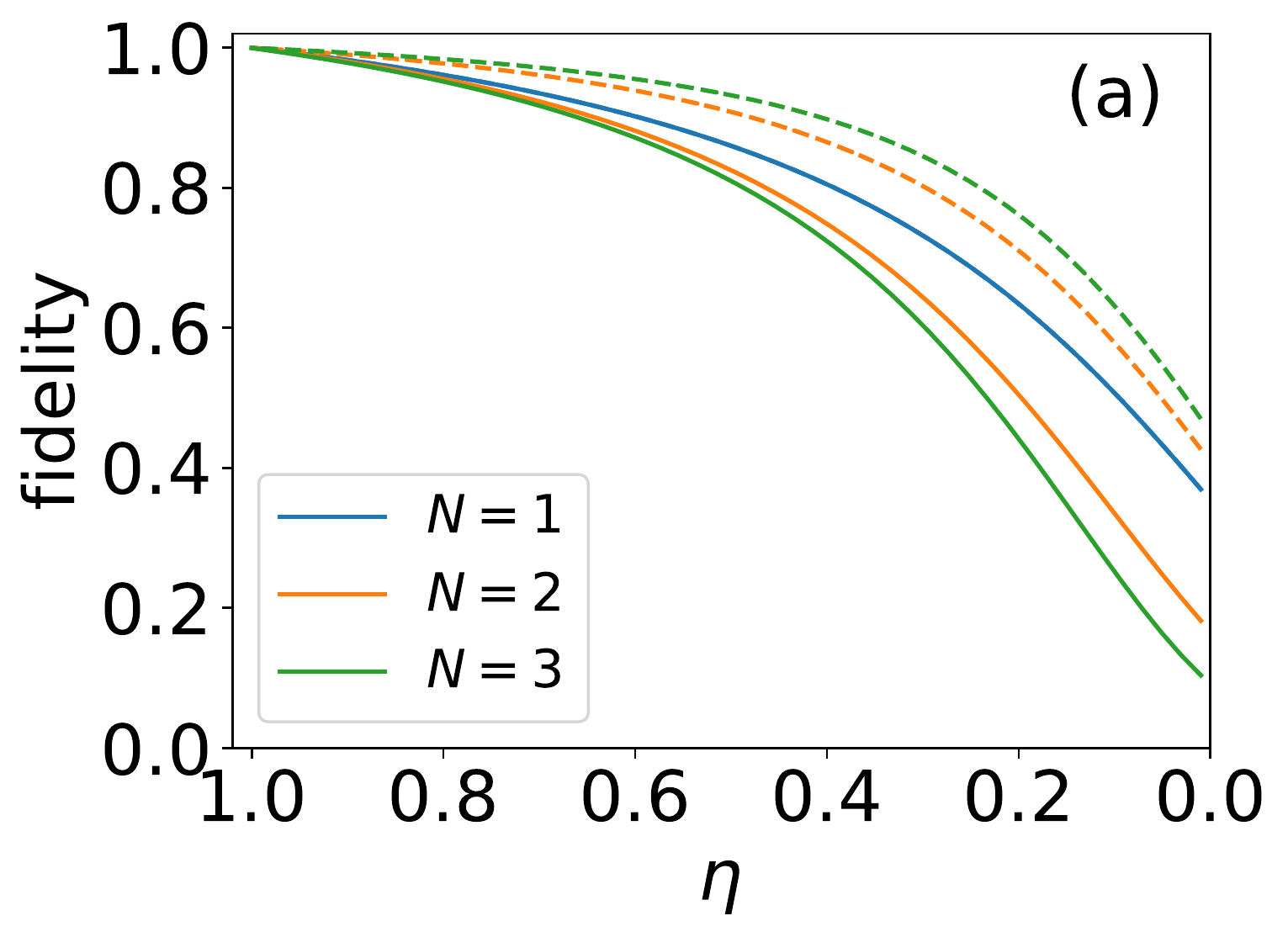}
	\includegraphics[width=0.49\columnwidth]{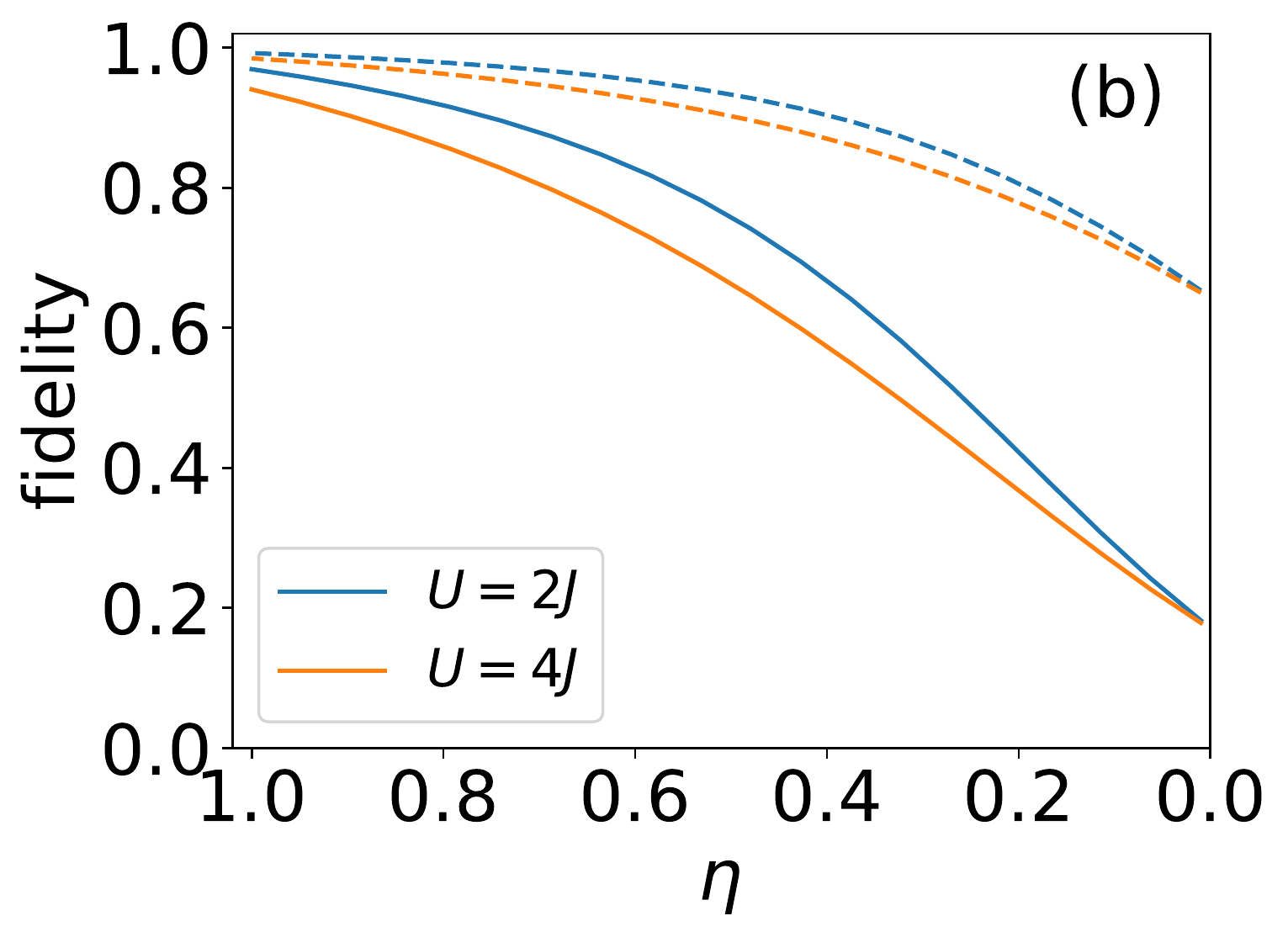}
	\caption{
		Steady-state fidelities $f$~(solid) and $\tilde f$~(dashed) as a function of the detection efficiency $\eta$ for (a) non-interacting case with different particle number $N$ at $M=8$ and for (b) interacting case with $M = N=4$ and $\gamma=0.01J$.}\label{eta}
\end{figure}

\textit{Imperfect detection efficiency.}---
So far, we have assumed perfect detection, $\eta=1$. For an efficiency $\eta<1$,
the ME becomes~\cite{PhysRevA.49.2133}
$\dot \rho = -i[H + H_{\rm fb}, \rho] + {\cal D}[A]\rho + \frac{1-\eta}{\eta}{\cal D}[F]\rho$.
In Fig.~\ref{eta}, we investigate the robustness against a reduction of $\eta$.
For both the non-interacting (a) and the interacting (b) cases,
the fidelity is found to decrease slowly with $\eta$. 
Roughly speaking, large fidelities require efficiencies larger than 50 percent.  

\begin{figure}
	\centering
	\includegraphics[width=0.99\columnwidth]{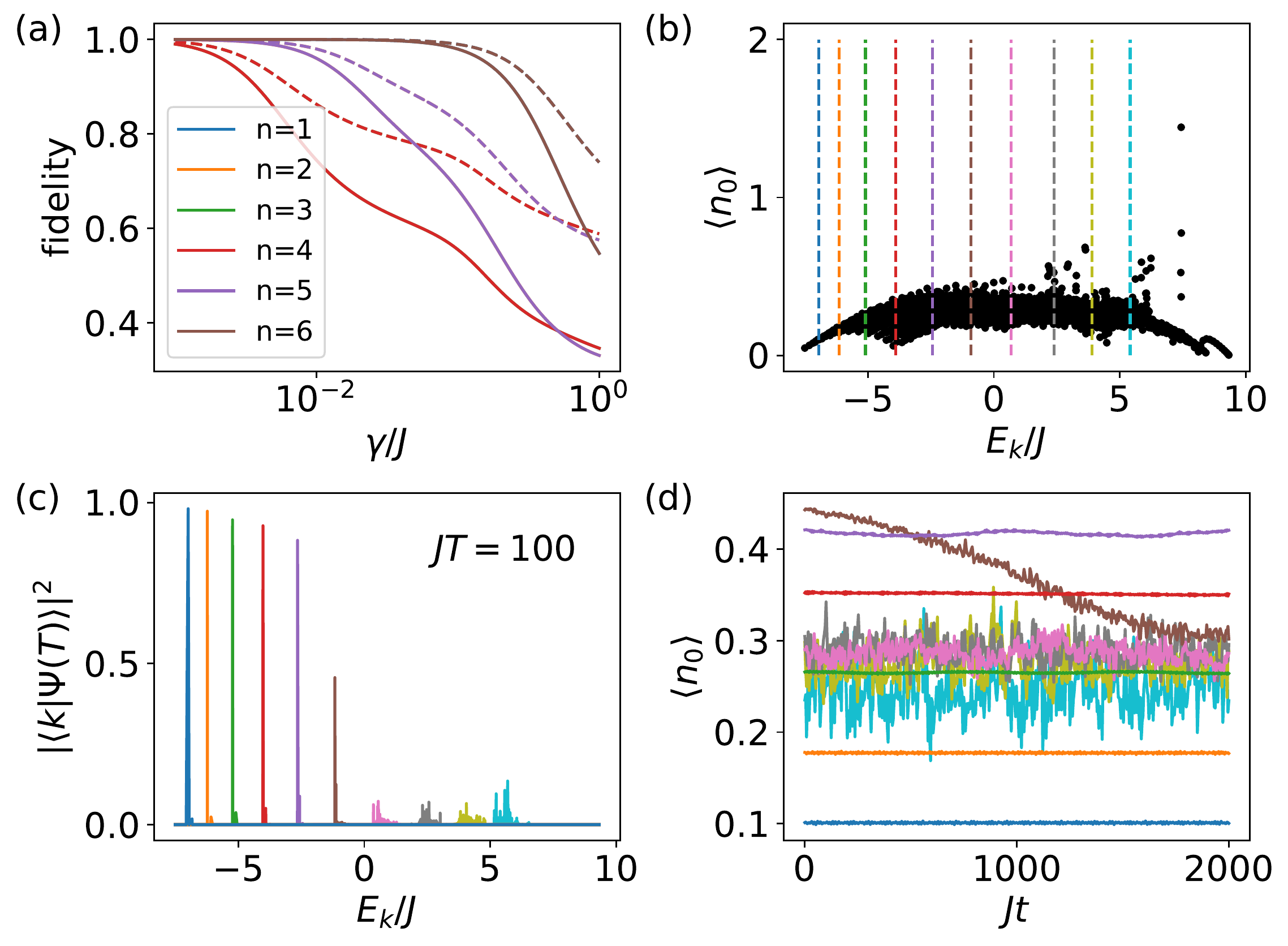}
	\caption{
		(a) Fidelities $f$~(solid) and $\tilde f$~(dashed) between the steady state of Eq.~\eqref{me_fbM} and the $n$th ``coherent'' eigenstate $|n,N\rangle$ versus $\gamma$ for $N=2$, $M=6$, $U=0$.  (b)~The expectation value of $n_0$ in the eigenstates at $U=J$. The dashed lines mark the states corresponding to $|n,N\rangle$. (c)~The distribution of the final state from adiabatic preparation (with the interaction strength $U$ linearly ramped up from $0$ to $J$ within $JT=100$ starting from the corresponding ``coherent'' eigenstate at $U=0$) on the eigenbasis at $U=J$. (d)~Evolution of $\langle n_0 \rangle$ under the Hamiltonian (1) with $U=J$ starting from the final state from adiabatic preparation. $N=4$ and $M=15$ in (b)-(d).}\label{eigen}
\end{figure}

\textit{Stabilizing excited eigenstates.}---
It is also possible to prepare excited eigenstates of the system. 
For the single-particle problem, the $n$th eigenstate
reads $|\psi_n\rangle =\sum\nolimits_{l}{g_l^{(n)}}|l\rangle$, with $g_l^{(n)} = \sqrt{\frac{2}{M+1}} \sin\big(\frac{n \pi l}{M+1}\big)$, with $n=1,2,\dots M$.
We can stabilize each of these eigenstates as steady state by using the feedback and measurement operators~\eqref{cF} with $z_l = [{g_{l+1}^{(n)} - g_{l-1}^{(n)}}]/{g_l^{(n)}}$.
This scheme can be generalized to $N$ non-interacting particles to prepare the $n$th ``coherent'' eigenstate $|n,N\rangle =\frac{1}{\sqrt{N!}}\big( \sum\nolimits_{l}{g_l^{(n)}}a_l^\dag\big)^N|0\rangle$, where all particles occupy the $n$th single-particle eigenstate.
Figure~\ref{eigen}(a) shows the fidelities between the steady state and $|n,N\rangle$ versus $\gamma$. By symmetry states $|n,N\rangle$ and $|M-n+1,N\rangle$ show the same behavior.
As expected, the fidelity increases and approaches $1$ as $\gamma$ decreases. The preparation of states closer to the center of the spectrum requires smaller $\gamma$.

The preparation of excited eigenstates of the interacting system offers the intriguing perspective to study their individual properties experimentally. This would allow, for instance, to directly probe ETH~\cite{Rigol2008, kaufman2016quantum, d2016quantum}, stating that each individual eigenstate of a generic ergodic system is characterized by expectation values that follow the prediction of the microcanonical ensemble. In Fig.~\ref{eigen}(b) we plot the mean occupations $\langle n_0\rangle$ of the leftmost site for all eigenstates of a system with $N=4$, $M=15$, and $U=J$ and one can see that the data points mostly form a rather thin stripe, confirming that (up to finite size effects) the expectation values depend mainly on the energy. Since it turns out that the stabilization of excited eigenstates via feedback control is difficult for the interacting system, we propose to use feedback control for the preparation of non-interacting eigenstates $|n,N\rangle$ and to then linearly switch on the interactions via a Feshbach resonance in a second step. In Fig.~\ref{eigen}(c), we plot the overlap of the time evolved states for a ramp time of $T=100/J$ with the interacting many-body eigenstates $|k\rangle$. The energy expected for perfect adiabatic behaviour is indicated by the vetical dashed lines of the same color in Fig.~~\ref{eigen}(b). While for initial coherent eigenstates with  smaller energy, we find a very narrow distribution with almost only a single eigenstate occupied, for larger energies broader distributions indicate non-adiabatic behaviour. We attribute the latter to the dynamic instability caused by the combination of repulsive interactions with negative effective mass found for states with large kinetic energy~\cite{MorschOberthaler2006}.
Even broader distributions are found~\cite{sm}, when preparing the state via a quench from a state with every other site occupied like in a recent experiment~\cite{Trotzky2012}. After the interaction ramp, we let the prepared state evolve with the interacting Hamiltonian and plot the mean-occupation versus time [Fig.~\ref{eigen}(d), colors like in (c)]. The fact that we hardly see any evolution of that expectation value for the states with the lower energies, shows how close they are to an eigenstate. In turn, for the prepared states in the upper half in the spectrum we find strong oscillations indicating the superposition of a few eigenstates.

\textit{Experimental implementation.}---
By including the feedback control term $I_{\rm hom} F$ to the Hamiltonian (\ref{BH}), the parameter for tunneling (rightwards) is modified to
$J' = J +i \sqrt{\gamma}I_{\rm hom} = \sqrt{J^2+\gamma I_{\rm hom}^2} e^{i\theta (t)}$, with
$\tan\theta(t) = \sqrt{\gamma}{I_{\rm hom}}/{J}$.  This can be achieved by accelerating the lattice potential $V_L(x)$ 
according to $V(x, t) = V_L(x - d_L\xi(t))$, with $d_L$ being the lattice constant.
In the reference frame co-moving with the lattice, the corresponding inertial force is described by a time-dependent vector potential represented by the time-dependent Peierls phase $\theta(t) = m d_L^2 \dot \xi(t)$ attached to the tunneling matrix element \cite{Eckardt2017}. Now choose $\dot \xi(t) = {\arctan(\sqrt{\gamma}{I_{\rm hom}}/{J})}/(m d_L^2) = \frac{2E_R}{\pi^2}{\arctan(\sqrt{\gamma}{I_{\rm hom}}/{J})}$ with $E_R=\pi^2/(2md_L^2)$. 

\begin{figure}
	\centering
	\includegraphics[width=1.\columnwidth]{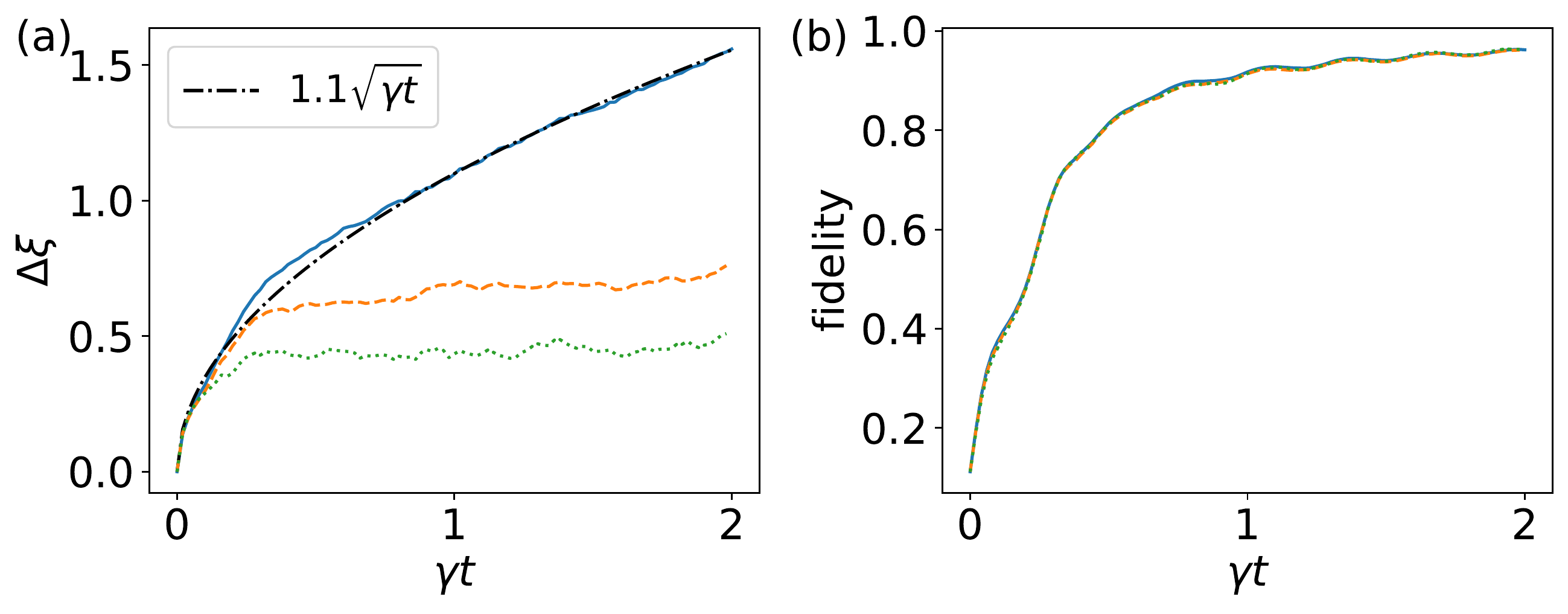}\\
	\caption{(a) Standard deviation $\Delta \xi$ of lattice drift $\xi$ versus time, averaged either over all 1000 trajectories of the SME~(solid) or over a postselected ensemble [dashed(dotted)], where the 50\%(85\%) of the trajectories were discarded that reach $\xi>1.8$ ($1.2$) during the evolution. (b) Corresponding fidelities between the instantaneous state and the ground state. The parameters are $N=2$, $M=4$, $U=J$, $dt=0.01/J$, $\gamma=0.1J$. Initially all atoms occupy the leftmost site.
	}\label{xi}
\end{figure}

The drift of lattice $\xi(t)$ is dependent on the homodyne current $I_{\rm hom}$,
and can be simulated from solving the modified SME. 
We show the standard deviation of the lattice displacement~(in unit of $d_L$), $\Delta \xi(t)$,
in Fig.~\ref{xi}(a) (blue solid line), with $E_R/J=15$ corresponding to a lattice depth of $V_0/E_R\approx5$~\cite{RevModPhys.80.885}.
$\Delta \xi(t)$ is found to be proportional to $\sqrt{\gamma t}$, a feature of a random walk.
At the time when the system approaches the steady state~[Fig.~\ref{xi}(b)],
the average drift of the lattice is about $1.5 d_L$.
The drift can be reduced further without loosing fidelity by postselecting those trajectories with $\xi(t)$ remaining below a given threshold (dashed and dotted lines).

\textit{Conclusion.}---We have proposed and investigated a simple scheme for measurement-based feedback control of a system of ultracold bosonic atoms in an optical lattice. Remarkably, at integer filling a simple modification of the feedback strength allows for cooling the system at arbitrary repulsive interaction strengths. Moreover, we showed that it is possible to prepare individual excited eigenstates, which would allow to experimentally study their statistical properties as they are predicted by the ETH. 
As an outlook, it will be interesting to generalize the scheme predicted here to scenarios, where \emph{in situ} cooling would be crucial, in order to counteract heating caused by Floquet engineering or imperfect adiabatic state preparation at a quantum phase transition, e.g. for the preparation of topological quantum states. Moreover, also the engineering of artificial thermal baths via feedback control, mimicking finite-temperature reservoirs and the combination of two of them to study heat-current-carrying steady states in a highly controllable quantum gas is a fascinating perspective. Generally, it will be interesting to employ feedback control to realize driven-dissipative systems and to control the properties of their non-equilibrium steady states (or phases) beyond the constraints of thermal equilibrium. 

\begin{acknowledgments}
	We acknowledge discussions with Lorenz Wanckel and Tobias Donner. This research was funded by the Deutsche Forschungsgemeinschaft (DFG, German Research
	Foundation) Projektnummer 163436311 SFB 910.
\end{acknowledgments}

\bibliography{ref}

\clearpage
\onecolumngrid
\begin{center}
  \textbf{\large Supplementary Material}\\[.2cm]
  Ling-Na Wu and Andr{\'e} Eckardt\\[.1cm]
  {\itshape Institut f\"ur Theoretische Physik, Technische Universit\"at Berlin,
Hardenbergstra\ss e 36, Berlin 10623, Germany}
\end{center}

\setcounter{equation}{0}
\setcounter{figure}{0}
\setcounter{table}{0}
\setcounter{page}{1}
\renewcommand{\theequation}{S\arabic{equation}}
\renewcommand{\thefigure}{S\arabic{figure}}
\renewcommand{\bibnumfmt}[1]{[S#1]}

\section{One particle in a double well}
For one particle in a double well, the system Hamiltonian reads
\begin{equation}
	H = \left( {\begin{array}{*{20}{c}}
			0&{ - J} \\
			{ - J}&0
	\end{array}} \right),
\end{equation}
whose ground state is given by $|G\rangle = \frac{1}{{\sqrt 2 }}\left( {\begin{array}{*{20}{c}}
		1 \\
		1
\end{array}} \right)$.
The feedback master equation (ME) reads
\begin{equation}\label{r2}
	\dot \rho = -i[H,\rho] + {\cal D}[A]\rho,
\end{equation}
with the collapse operator given by
\begin{equation}
	A = {{\sqrt \gamma  }}\left( {{\sigma _z} - i{\sigma _y}} \right) = {{\sqrt \gamma  }}\left( {\begin{array}{*{20}{c}}
			1&{ - 1} \\
			1&{ - 1}
	\end{array}} \right).
\end{equation}
Note that the additional feedback-induced term in the Hamiltonian $H_{\rm fb} \propto \{\sigma_y, \sigma_z\} = 0 $ in this case.

Starting from the initial state $\left( {\begin{array}{*{20}{c}}
		1 \\
		0
\end{array}} \right)$, the state at time $t$ is given by
\begin{equation}
	\rho \left( t \right) = \frac{1}{2}\left( {\begin{array}{*{20}{c}}
			{1 + {e^{ - 2\gamma t}}\cos (2Jt)}&{1 - {e^{ - 4\gamma t}} - i{e^{ - 2\gamma t}}\sin (2Jt)} \\
			{1 - {e^{ - 4\gamma t}} + i{e^{ - 2\gamma t}}\sin (2Jt)}&{1 - {e^{ - 2\gamma t}}\cos (2Jt)}
	\end{array}} \right).
\end{equation}
Hence, we have
\begin{equation}
	\langle G|\rho \left( t \right)|G\rangle  = 1 - \frac{1}{2}{e^{ - 4\gamma t}}.
\end{equation}

\section{The nonunique steady state for odd site number~($M$) case}
For the non-interacting case, when the site number $M$ is odd, the steady state of the system is not unique.
This is attributed to the presence of more than one dark state of the collapse operator $A$ due to the degeneracy of the system.
To better understand it, we can take a look at an example with $M=3$ and $N=2$. There are three single-particle eigenstates
in this case, with energies $E_{-1}=-\sqrt{2}J, E_0=0, E_{1}=\sqrt{2}J$. Therefore, the system has two degenerate eigenstates with eigenenergy 
$0$, i.e., one state with two particles in state $|E_0\rangle$, and one state with one particle in $|E_1\rangle$ and the other in $|E_{-1}\rangle$.
When expressed in the Fock state representation of the eigenstate basis, $|n_{-1}, n_{0}, n_{-1}\rangle$, these two states read
$|0,2,0\rangle$ and $|1,0,1\rangle$. Any superposition of these two states is also an eigenstate of the system,
$|\psi\rangle = \alpha |0,2,0\rangle + \beta |1,0,1\rangle$. Both components will be pumped to the state
$|1,1,0\rangle$ by the collapse operator $A$. A proper choice of the coefficients $\alpha$ and $\beta$
will then lead to $A|\psi\rangle=0$. Namely, besides the ground state $|2,0,0\rangle$, there is another
dark state of $A$. Therefore, if the initial state has an overlap with this dark state, this component will not be influenced 
by the action of $A$ and remain the same during the evolution. In this case, the steady state won't be the ground state of the system.
The key point in the above reasoning is the degeneracy of the system, which leads to the appearance of other
dark states~(besides the ground state) due to the `destructive' interference between the degenerate eigenstates.
When such a degeneracy is lifted, for instance, by including interactions, the steady state becomes unique~(the ground state).
When the site number is even, it can be shown that the ground state is the only dark state of $A$ in spite of the degeneracy of the system.

\section{Energy difference between the steady state and the ground state}
\begin{figure}[H]
	\centering
	\includegraphics[width=0.8\columnwidth]{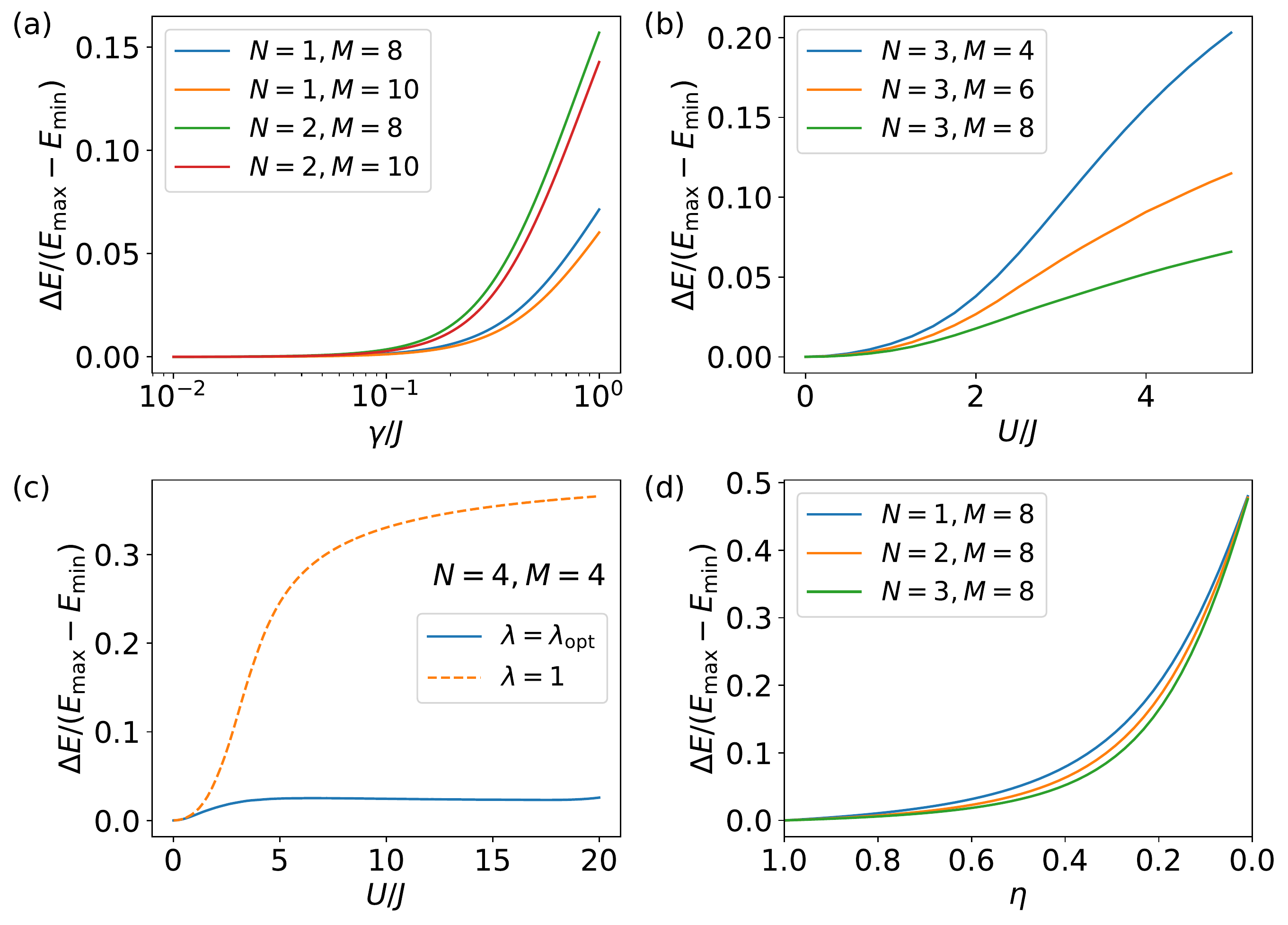}\\
	\caption{The energy difference between the steady state and the ground state $\Delta E$ normalized by the width of the spectrum $(E_{\rm max}-E_{\rm min})$, as a function of (a) the measurement strength $\gamma$ with $U=0$, (b)-(c) the interaction strength $U$, and (d) the detection efficiency $\eta$ with $U=0$. The measurement strength in (b)-(d) is $\gamma=0.01J$. In (a), (b), and (d), $\lambda=1$. (c) compares the results with $\lambda=1$ and the optimized $\lambda_{\rm opt}$.}\label{energy}
\end{figure}

As a second measure of the effectiveness of our cooling scheme, we calculate the energy difference between the steady state and the ground state $\Delta E$, normalized by the width of the spectrum $(E_{\rm max}-E_{\rm min})$.
The results are shown in Fig.~\ref{energy}.

\section{External trapping potential}
\begin{figure}[H]
	\centering
	\includegraphics[width=0.8\columnwidth]{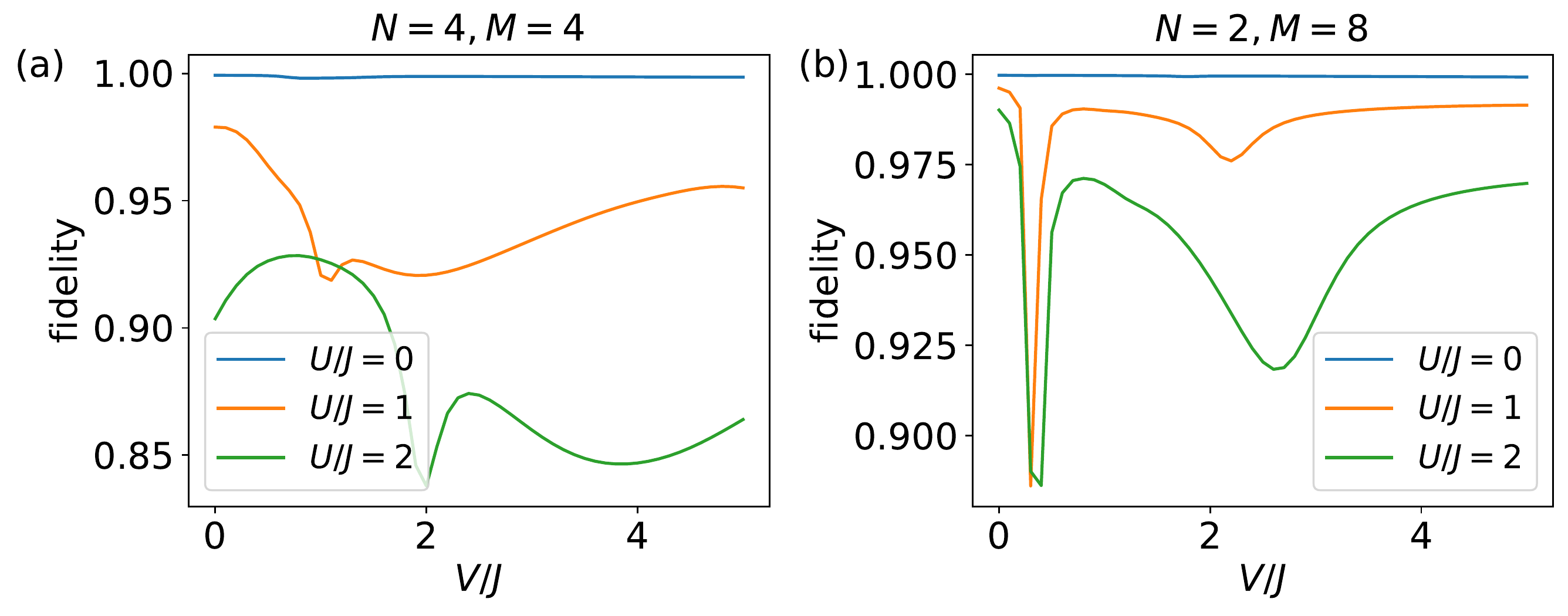}
	\caption{The fidelity between the steady state and the ground state as a function of the potential strength $V$.}\label{potential}
\end{figure}
Here we consider the case when there is an external harmonic trapping potential
\begin{eqnarray}
	H_V = \frac{V}{2} \sum\limits_{l=1}^{M}{(l-l_0)^2 n_l},
\end{eqnarray}
with $l_0=(M-1)/2$. To cool the system in this case, one should use measurement operator $c=\sum\nolimits_l {{z_l} n_l}$ with $z_l = [{g_{l+1} - g_{l-1}}]/{g_l}$ and $g_l$ being the coefficient of the ground state for the system with the on-site potential. Figure~\ref{potential} 
shows the fidelity between the steady state and the ground state as a function of the potential strength $V$. We can see that for the non-interacting systems, a close-to-one fidelity can be obtained no matter how strong the potential is. For the interacting systems, the fidelity is degraded due to the potential, but the influence is found to be weak.

\section{The robustness of the free particle scheme against disorder}
\begin{figure}[H]
\centering
\includegraphics[width=0.4\columnwidth]{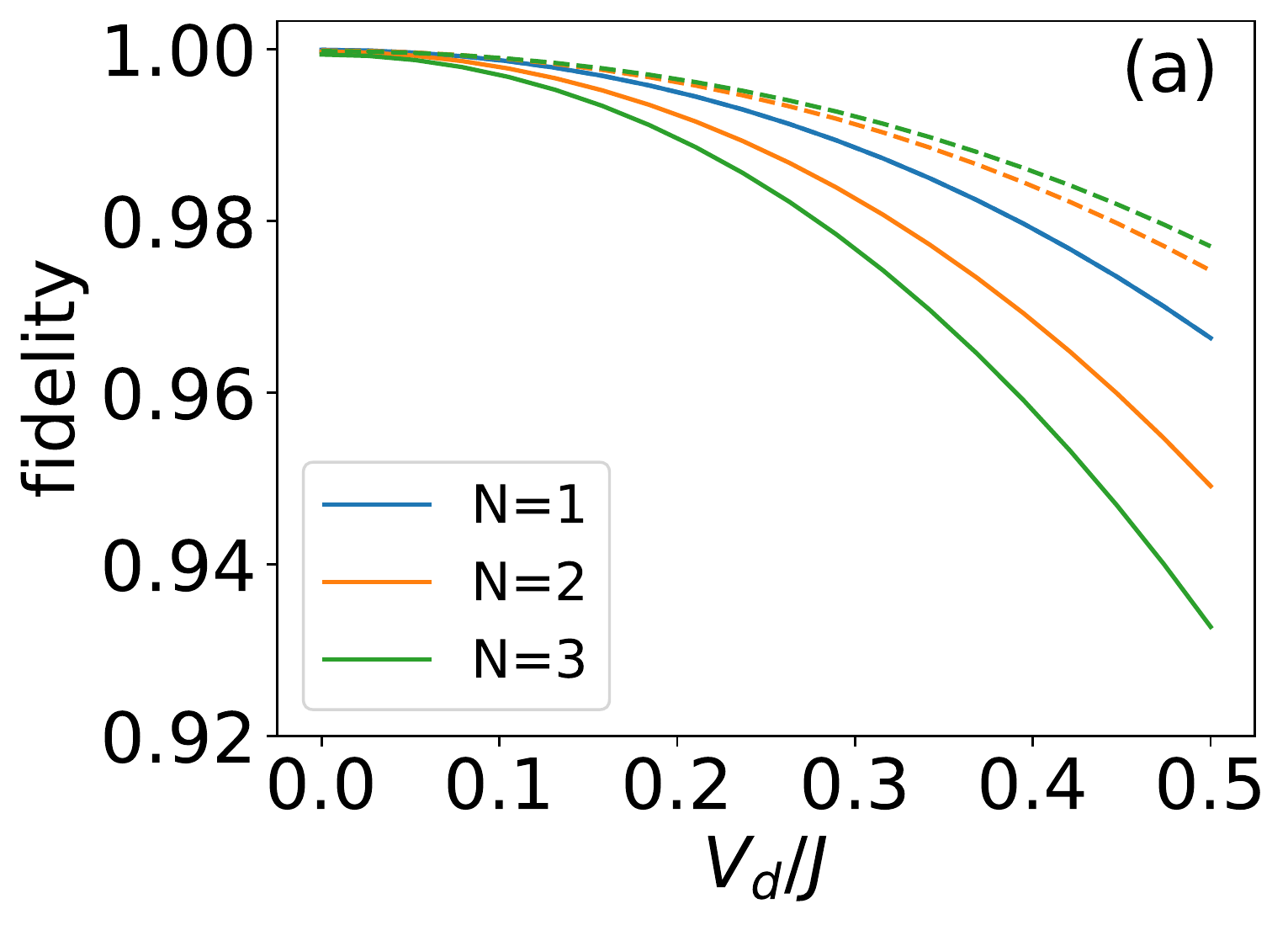}
\includegraphics[width=0.4\columnwidth]{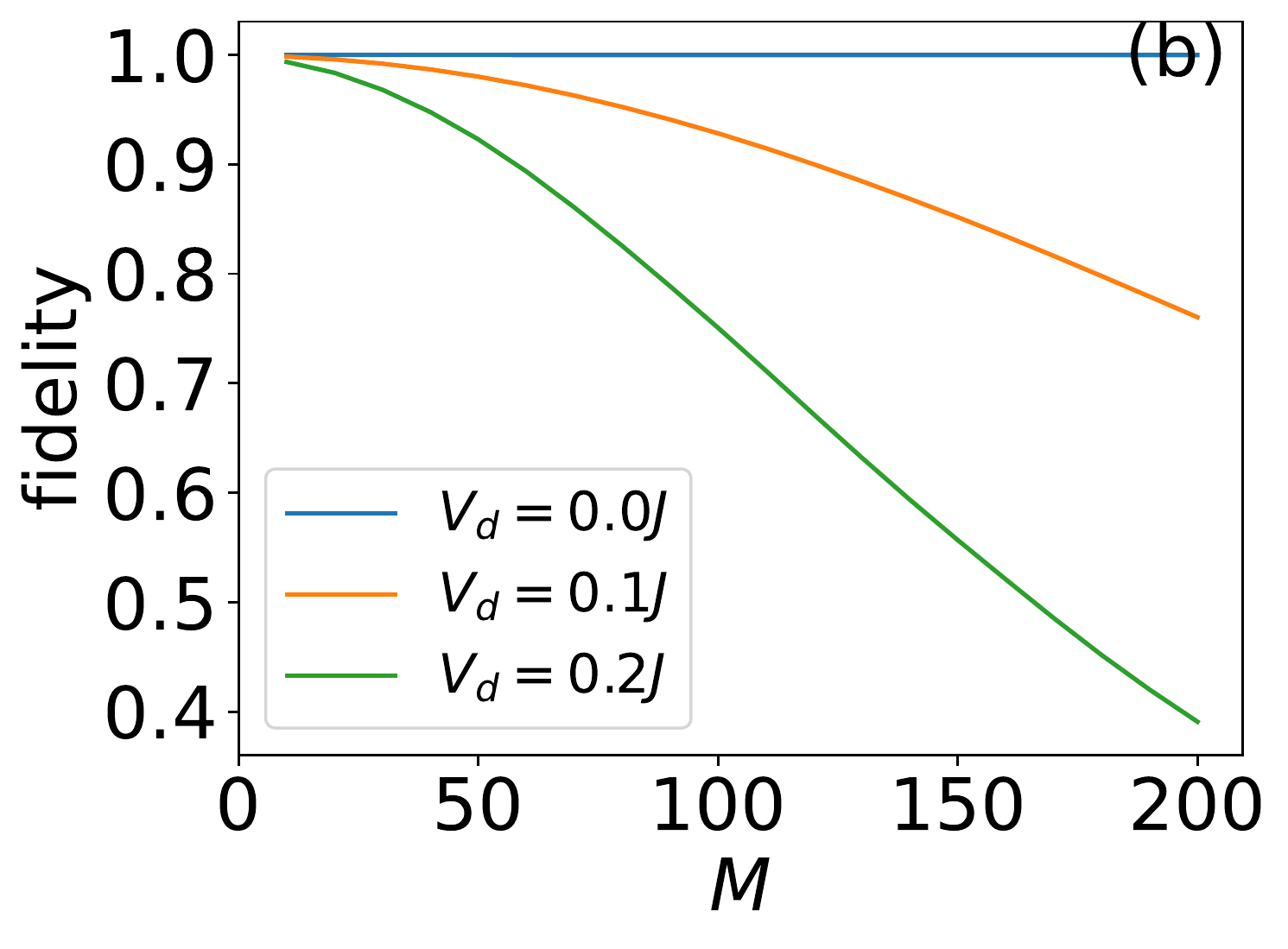}
\caption{
	The fidelities $f$~(solid) and $\tilde f$~(dashed) between the steady state of the proposed scheme and the ground state of Eq.~\eqref{Hd} as a function of (a) the disorder strength $V_d$ for different $N$ at $M=8$, and (b) the system size $M$ at $N=1$.  Other parameters are $U=0$, $\gamma=0.01J$.}\label{disorder}
\end{figure}

As a test of the robustness of the free particle scheme,
we add quasi disorder via an incommensurate periodic potential modulation~(Aubry-Andr{\'e} model~\cite{AA1980})
$V(l)=V_d\cos(2\pi\beta l)$, with $\beta=(\sqrt{5}-1)/2$, i.e., consider the Hamiltonian
\begin{equation}\label{Hd}
H = -J\sum\nolimits_{l}(a_l^\dag a_{l+1} + a_{l+1}^\dag a_l) + \sum\nolimits_{l} V(l) n_l,
\end{equation}
and see how the fidelity decreases with the disorder strength $V_d$.
The results are shown in Fig.~\ref{disorder}, where
$V_d$ remains below the critical value of $V_d^c = 2J$,
so that all eigenstates remain delocalized.
We can see that the scheme is robust to weak disorder.

\section{System size dependence of the measurement strength $\gamma$}
\begin{figure}[H]
\centering
\includegraphics[width=0.8\columnwidth]{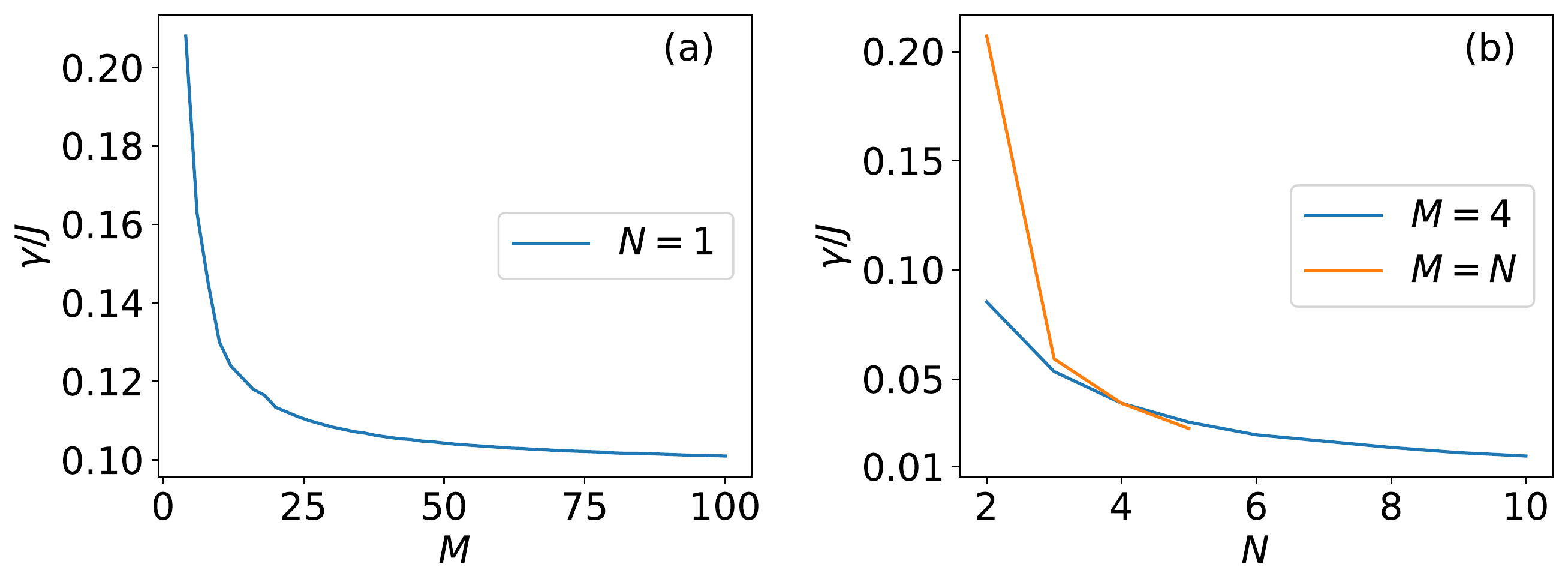}\\
\caption{The maximal measurement strength $\gamma$ needed to reach a fidelity of $0.99$ as a function of (a) $M$ with $N=1$, and (b) $N$ with $M=4$ or $M=N$. The interaction strength is $U=0$.}\label{gamma}
\end{figure}
Here we consider the system size dependence of the measurement strength $\gamma$.
The fidelity increases as the measurement strength $\gamma$ decreases. Figure~\ref{gamma} shows the maximal $\gamma$ to reach a fidelity of $0.99$ for various system sizes. From (a), one can see that the dependence of $\gamma$ on $M$ is weak when $N$ is fixed. As for the dependence on $N$ [see (b)], $\gamma$ shows a fast decay with increasing $N$ at the beginning and then the decrease becomes slow as $N$ increases further.

\section{Double well system under weak interactions}
According to perturbation theory, for weak interactions $NU/2 \ll J$,
the ground state is approximately given by
$|G\rangle \simeq |\psi_0^{(0)}\rangle - \alpha |\psi_2^{(0)}\rangle$, where
$|\psi_n^{(0)}\rangle$ denotes the $n$th eigenstate of the non-interacting Hamiltonian and
$\alpha=\sqrt{2N(N-1)}(U/J)/16$.
~{{By noting $J_-|\psi_2^{(0)}\rangle=\sqrt{2(N-1)}|\psi_1^{(0)}\rangle$ and $J_-|\psi_1^{(0)}\rangle=\sqrt{N}|\psi_0^{(0)}\rangle$,
	we see that the ground state is coupled to the first excited state $|E\rangle \simeq |\psi_{1}^{(0)}\rangle$ by the collapse operator $A=\sqrt{\gamma}J_-$ and the ratio of transfer rate between them reads
	$|\langle E|J_-|G\rangle/\langle G|J_-|E\rangle|^2 = [\sqrt{2(N-1)}\alpha/\sqrt{N}]^2=[(N-1)(U/J)/8]^2 \equiv \nu$,
	the fidelity is approximately given by $f^2=1-\nu$.
	For general $M$, the scenario is more complicated as the ground state is coupled to several states,
	while the conclusion is generally that the reduction of the fidelity is proportional to $[nU/(2\Delta E)]^2$
	with the filling factor $n=N/M$ and single-particle energy gap $\Delta E=2J\cos[\pi/(M+1)]-2J\cos[2\pi/(M+1)]$.}}

\section{Double well system with odd particle numbers}

For odd $N$ and $M=2$, when $U \gg NJ$, the ground state
is approximately given by
\begin{equation}
|G\rangle \simeq \frac{1}{\sqrt{2}}\left(|\frac{N+1}{2},\frac{N-1}{2}\rangle + |\frac{N-1}{2},\frac{N+1}{2}\rangle\right) + \frac{NJ}{4\sqrt{2}U}\left(|\frac{N+3}{2},\frac{N-3}{2}\rangle + |\frac{N-3}{2},\frac{N+3}{2}\rangle\right).
\end{equation}
Therefore, for $N \gg 1$ we have
\begin{eqnarray}
\frac{1}{\sqrt{\gamma}}A|G\rangle &=& [n_1-n_2-\lambda(a_1^\dag a_2 - a_2^\dag a_1)]|G\rangle \notag\\
&\simeq& \frac{1}{\sqrt{2}}\left(|\frac{N+1}{2},\frac{N-1}{2}\rangle - |\frac{N-1}{2},\frac{N+1}{2}\rangle\right) + \frac{3NJ}{4\sqrt{2}U}\left(|\frac{N+3}{2},\frac{N-3}{2}\rangle - |\frac{N-3}{2},\frac{N+3}{2}\rangle\right) \notag\\
&& -\frac{\lambda N}{2\sqrt{2}}\left(|\frac{N+3}{2},\frac{N-3}{2}\rangle + |\frac{N+1}{2},\frac{N-1}{2}\rangle\right) - \frac{\lambda N^2J}{8\sqrt{2}U}\left(|\frac{N+5}{2},\frac{N-5}{2}\rangle + |\frac{N-1}{2},\frac{N+1}{2}\rangle\right) \notag\\
&& +\frac{\lambda N}{2\sqrt{2}}\left(|\frac{N-1}{2},\frac{N+1}{2}\rangle + |\frac{N-3}{2},\frac{N+3}{2}\rangle\right) + \frac{\lambda N^2J}{8\sqrt{2}U}\left(|\frac{N+1}{2},\frac{N-1}{2}\rangle + |\frac{N-5}{2},\frac{N+5}{2}\rangle\right)\notag\\
&=& |G\rangle + \frac{\lambda N}{2\sqrt{2}}\left(\frac{ N J}{4U}-1\right)\left(|\frac{N+1}{2},\frac{N-1}{2}\rangle - |\frac{N-1}{2},\frac{N+1}{2}\rangle\right) \notag\\
&& + \frac{N}{2\sqrt{2}}\left(\frac{J}{U}-\lambda\right)\left(|\frac{N+3}{2},\frac{N-3}{2}\rangle - |\frac{N-3}{2},\frac{N+3}{2}\rangle\right) \notag\\
&& -\frac{\lambda N^2J}{8\sqrt{2}U}\left(|\frac{N+5}{2},\frac{N-5}{2}\rangle - |\frac{N-5}{2},\frac{N+5}{2}\rangle\right).
\end{eqnarray}
From above expression, we can see that it's impossible to make $A|G\rangle = 0$ no matter what value $\lambda$ takes.
We can only make the transfer rate from the ground state to other states small by setting $\lambda=J/U$.

\section{Multi-site case (general $M$) at integer filling}

Now let's consider the general $M$ case with integer filling factor $n=N/M$.
For $U \gg NJ$, the ground state is well approximated by
\begin{equation}
|G\rangle \simeq |n,\ldots, n\rangle +  \frac{n J}{U} \sum\limits_{l=1}^{M-1}{(|n,\ldots,n_l+1,n_{l+1}-1,\ldots,n\rangle + |n,\ldots,n_l-1,n_{l+1}+1,\ldots,n\rangle)},
\end{equation}
where the second term contains all the Fock states that can be obtained from the equally distributed state $|n,\ldots, n\rangle$
by transferring one particle from one site to its neighboring site.
This gives
\begin{equation}
\frac{1}{\sqrt{\gamma}}c|G\rangle = \sum\limits_{l=1}^{M}{z_l n_l}|G\rangle \simeq \frac{n J}{U} \sum\limits_{l=1}^{M-1}{(z_l-z_{l+1})(|n,\ldots,n_l+1,n_{l+1}-1,\ldots,n\rangle - |n,\ldots,n_l-1,n_{l+1}+1,\ldots,n\rangle)},
\end{equation}
where we have used $\sum\nolimits_{l} z_l=0$, and
\begin{equation}
\frac{i}{\sqrt{\gamma}}F|G\rangle = \sum\limits_{l=l}^{M-1}(a_l^\dag a_{l+1} - a_{l+1}^\dag a_l)|G\rangle \simeq n \sum\limits_{l=1}^{M-1}{(|n,\ldots,n_l+1,n_{l+1}-1,\ldots,n\rangle - |n,\ldots,n_l-1,n_{l+1}+1,\ldots,n\rangle)}.
\end{equation}
Therefore,
\begin{equation}
\frac{1}{\sqrt{\gamma}}A|G\rangle = \frac{1}{\sqrt{\gamma}}(c-i\lambda F)|G\rangle  \simeq n \sum\limits_{l=1}^{M-1}{\left[(z_l-z_{l+1})\frac{J}{U}-\lambda\right](|n,\ldots,n_l+1,n_{l+1}-1,\ldots,n\rangle - |n,\ldots,n_l-1,n_{l+1}+1,\ldots,n\rangle)}.
\end{equation}
So by setting $\lambda = (z_l-z_{l+1})\frac{J}{U}$, we have $A|G\rangle \simeq 0$.

\section{State distribution in the eigenbasis}
\begin{figure}[H]
\centering
\includegraphics[width=0.8\columnwidth]{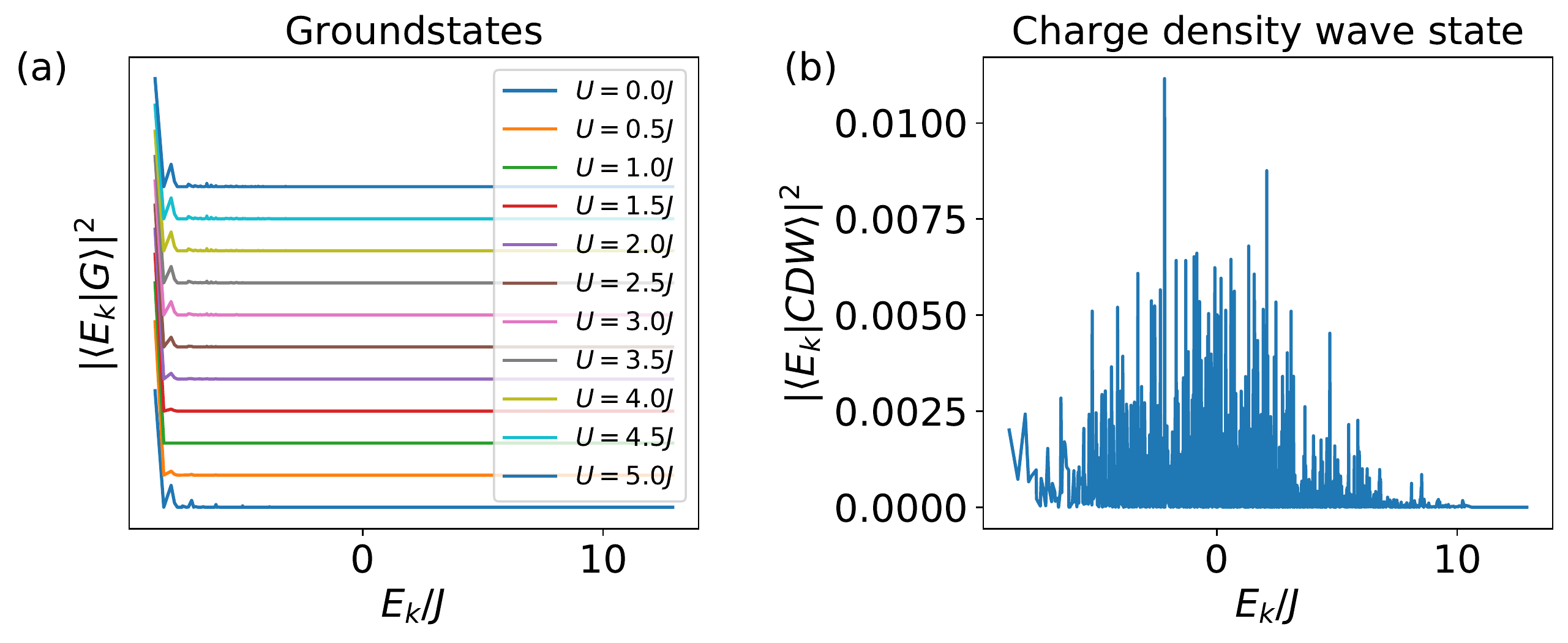}
\caption{The distribution of some states in the eigenbasis at $U=J$. (a) The ground states at various interaction strength $U$. (b) The charge density wave state with every second site occupied. The system sizes are $N=5$ and $M=10$.}\label{quench}
\end{figure}

Figure~\ref{quench} shows the distribution of some states, which can be prepared experimentally, in the eigenbasis at $U=J$.
(a) shows the ground states at various interaction strength $U$. One can see that all of them
have a large overlap with the low energy states.
(b) shows the state with every second site occupied. This state has a broad distribution with a center in the middle of the spectrum.

\end{document}